\documentclass[osajnl2,showpacs,twocolumn]{revtex4}

\begin{document}

\title{Rigorous analysis of bistable memory in silica toroid microcavity}

\author{Wataru Yoshiki$^{1}$ and Takasumi Tanabe$^{1,*}$}
\address{$^1$Department of Electronics and Electrical Engineering, Faculty of Science and Technology, Keio University, \\ 3-14-1, Hiyoshi, Kohoku-ku, Yokohama 223-8522, Japan}
\address{$^*$Corresponding author: takasumi@elec.keio.ac.jp}

\begin{abstract}
We model the nonlinear response of a silica toroid microcavity using coupled mode theory and a finite element method, and successfully obtain Kerr bistable operation that does not suffer from thermo-optic effect by optimizing the fiber-cavity coupling.  
Our analysis shows it is possible to demonstrate a Kerr bistable memory with a memory holding time of $500~\mathrm{ns}$ at an extremely low energy consumption.
\end{abstract}

\maketitle

\section{Introduction}
Optical bistability is a fundamental physical phenomenon where it is possible for certain devices to have two stable transmission states.  
It occurs when the refractive index or the absorption of the nonlinear medium in an optical cavity is dependent on the light intensity.  
Optical bistable devices are considered to be an important building blocks in all-optical signal processing for such components as optical memories and optical flip-flops \cite{tsuda1990a}, and this phenomenon was extensively studied in 1970s to 80s \cite{gibbs1985a}.  
However, at that time, the size of the cavity and the operating energy were too large to be considered for practical applications.

Recent progresses in achieving higher quality factors ($Q$) in ultrasmall microcavities on-chip \cite{noda2003a, tanabe2007a, almani2003a, vahala2003a} has refocused attention on optical bistability \cite{soljacic2002a}, due to the possibility of achieving denser integration and lower energy consumption.  
Since the photon density in a cavity scales with $Q/V$, where $V$ is the mode volume, a high $Q$ cavity with a small $V$ enables us to use various nonlinearities at extremely low input powers;  hence it allows optical bistability at an ultralow power.

Solja\v{c}i\'c \textit{et al.} demonstrated numerically that optical bistability based on Kerr nonlinearity is possible by using a two-dimensional photonic crystal nanocavity at an ultralow driving power of 133~mW \cite{soljacic2002a}.  
Various experiments have already been reported in silica microspheres \cite{collot1993a}, silicon photonic crystals \cite{notomi2005a, notomi2009a}, and silicon microring resonators \cite{almeida2004b}, but by using the thermo-optic (TO) effect.  
Since the TO effect is accompanied by thermal accumulation, the response is relatively slow.  
To achieve faster speed, the carrier-plasma effect has been utilized.  
This effect is the result of carrier generation \cite{tanabe2005a, shinya2008a}.  
The latest research has reported on 4-bit optical random-access memory operation based on carrier nonlinearity, at a power consumption of only 30~nW, by using InGaAsP photonic crystal nanocavities \cite{nozaki2012a}.  
The key to achieving the low power consumption is the smallness of the cavity $V$.

As introduced above, the required power for the operation of optical bistability has been significantly reduced in recent years due to the high $Q/V$.  
However, all of the demonstrations use either TO or carrier-plasma effects to drive the bistability whose nonlinearities are accompanied by photon absorption.  
Recent advances on linear and nonlinear studies of microcavities and chip-based waveguide devices have opened possibility of their use not only for classical all-optical processing but also for loss sensitive applications such as quantum information processing \cite{spillane2005a, yoshie2004a, politi2008a, takesue2007a}.  
With those applications in mind, we need to reduce significantly  the power loss (consumption) of the bistable system.  
The use of the optical Kerr effect is the ultimate goal because it does not absorb photons.  
In addition, the use of an ultra-high $Q$ allows us to reduce the power scattering loss of the signal light from the system, and this will support all-optical information processing for loss-sensitive applications.

Although there have been several attempts to employ the optical Kerr effect by using large-bandgap materials such as AlGaAs \cite{kawashima2008a} and chalcogenide glasses \cite{eggleton2011a}, it still seems to be difficult to obtain Kerr bistability without suffering from the carrier effect or TO effect \cite{delong1989a, kang1994a, collot1993a}.  
On the other hand, silica has been an excellent material with which to study of various aspects of $\chi^{(3)}$ based physics \cite{agrawal1995a}, because of the large bandgap that can suppress carrier generation.  
Therefore silica microcavities have the potential to achieve a Kerr bistable memory that consumes very little energy.

Among various silica microcavities, the silica toroid microcavity \cite{almani2003a} has an ultra-high $Q$ and is capable of integration on a chip.  
As discussed above, a high $Q$ cavity is attractive for both achieving optical Kerr bistability and for low-loss applications.

In this paper, we demonstrate numerically that an optical bistable memory based on the optical Kerr effect is possible by controlling the coupling between the cavity and the tapered fiber.  
Our model consists of coupled mode theory (CMT) and the finite element method (FEM).  

The paper is organized as follows.  
First in \S\ref{sec1}, we provide a simple picture of the physics that we are going to employ. 
In \S\ref{sec2} we describe a numerical simulation model that combines CMT and FEM.  
In \S 4 we show the calculation result and in \S 5 we discuss the energy consumption.  
Finally, we finish with a conclusion.  

\section{Simple model}\label{sec1}
As described in the previous section, it is still difficult to use the optical Kerr effect in ultrahigh-$Q$ silica microcavities, even though the material has a large bandgap.  
This is because of the small light absorption at the surface, which is caused by a thin water layer \cite{collot1993a,rokhsari2004a}.  
Therefore, we need to analyze this carefully by using rigorous modeling and applying realistic physical parameters to reveal the conditions required for obtaining Kerr bistability.  
However, before undertaking a rigorous analysis we start with a simple model to gain an intuitive understanding of the strategy for obtaining Kerr bistability.  
In this section we pursue an analytical discussion about how we can obtain the Kerr effect without exhibiting considerable TO effect.

First we derive the relationship between the wavelength shift and the input energy required for Kerr nonlinearity.  The energy $U$ of an electromagnetic wave in a dielectric medium is given by \cite{yariv2010a},
\begin{eqnarray}
 U = \frac{1}{4}\varepsilon E_0^2 V,
\label{eq:index3}
\end{eqnarray}
where $\varepsilon$, $E_0$ and $V$  are the dielectric constant, the electric field amplitude and the mode volume, respectively.  Using Eq.~(\ref{eq:index3}) and $I = v\varepsilon |E_0|^2/2$, where $v$ is the light speed of the medium, we obtain the relation between the power density $I$ and the optical energy $U$ as,
\begin{eqnarray}
I = \frac{2c}{n_0} \frac{U}{V},
\label{eq:index4}
\end{eqnarray}
where $c$ is the velocity of light and $n_0$ is the refractive index of the cavity medium.  Then, the refractive index change caused by the Kerr effect is given by,
\begin{eqnarray}
\Delta n_{\mathrm{Kerr}} = n_2 I =  \frac{2 n_2 c}{n_0} \frac{U}{V},  
\label{eq:kerr9}
\end{eqnarray}
where $n_2$ is the nonlinear refractive index.  
Equation~(\ref{eq:kerr9}) allows us to calculate the refractive index change as a function of the energy in the cavity.  

Next we discuss the TO nonlinearity.
The refractive index change caused by the TO effect $\Delta n_{\mathrm{TO}}$ is given by,
\begin{eqnarray}
\Delta n_{\mathrm{TO}} = n_0 \xi T.
\label{eq:index51}
\end{eqnarray}
Where $\xi = (1/n) (\partial n / \partial T)$ is the TO coefficient.  
The energy required to increase the temperature of 1~K for volume $V$ is given by $C\rho V$, where $C$ is the heat capacity and $\rho$ is the density of the material.  
Then, using Eq.~(\ref{eq:index51}), we obtain
\begin{eqnarray}
\Delta n_{\mathrm{TO}} = \frac{n_0 \xi}{C \rho}\frac{1}{V}U_{\mathrm{abs}},
\label{eq:index52}
\end{eqnarray}
where $U_{\mathrm{abs}}$ is the energy absorbed by the material.  
Equations~(\ref{eq:kerr9}) and (\ref{eq:index52}) describe the refractive index change at a given energy for the Kerr and TO effects respectively.  
For the Kerr effect to be larger than the TO effect the following condition is required.
\begin{eqnarray}
\frac{\Delta n_{\mathrm{Kerr}}}{\Delta n_{\mathrm{TO}}} = \frac{2n_2 c \rho C}{n_0^2 \xi} \frac{U}{U_{\mathrm{abs}}} > 1
 \label{eq:eq8}
\end{eqnarray}
To gain a simple understanding, we assume a steady-state model.  
The light energy in the cavity is constant and generates heat at a constant rate.  Thus, $U(t)$ and $U_\mathrm{abs}(t)$ are expressed as,
\begin{eqnarray}
\left\{
\begin{array}{lll}
U(t) & = & U_0 \\
\frac{\mathrm{d}U_\mathrm{abs}(t)}{\mathrm{d}t} & = & - \frac{1}{\tau_\mathrm{diff}}U_\mathrm{abs} + \frac{1}{\tau_\mathrm{abs}}U(t) ,
 \label{eq:eq9}
\end{array}
\right.
\end{eqnarray}
where $U_0$, $\tau_\mathrm{diff}$ and $\tau_\mathrm{abs}$ are the steady energy in the cavity, the thermal relaxation time and the thermal generation rate caused by photon absorption, respectively.  
When we neglect the thermal diffusion ($\tau_\mathrm{diff}^{-1} = 0$), Eq.~(\ref{eq:eq8}) is simply expressed as,
\begin{eqnarray}
 t< \frac{2n_2 c \rho C}{n_0^2 \xi} \tau_{\mathrm{abs}}.
\label{eq:eq12}
\end{eqnarray}
This provides direct view as to how we should design the cavity system in order to obtain the Kerr effect without the TO effect being too great.  
When we use the following parameters for $\mathrm{SiO_2}$, $\xi=5.2\times 10^{-6}~\mathrm{K^{-1}}$, $n_2=3.67\times 10^{-20}~\mathrm{m^2/W}$ $n=1.47$, $C=7.41\times 10^{-1}~\mathrm{J/(g\cdot K)}$, and $\rho=2.65~\mathrm{g/cm^3}$, Eq.~(\ref{eq:eq12}) gives the following condition,
\begin{eqnarray}
t < 3.84 \tau_{\mathrm{abs}}.
 \label{eq:eq13}
\end{eqnarray}
This equation provides a simple understanding of how we can achieve Kerr nonlinearity in $\mathrm{SiO_2}$ microcavities.  
Although the Kerr effect governs the refractive index change ($\Delta n_{\mathrm{Kerr}} > \Delta n_{\mathrm{TO}}$) at an early stage of the operation, the TO effect becomes dominant after a period of time given by Eq.~(\ref{eq:eq13}).  
When we employ a $\tau_\mathrm{abs}$ of $329~\mathrm{ns}$ (the selection of this value is discussed in detail in \S\ref{sec:photon}), Eq.~(\ref{eq:eq13}) gives $t<1.26~\mathrm{\mu s}$, which is the duration time, for which $\Delta n_\mathrm{Kerr}$ is larger than $\Delta n_\mathrm{TO}$.  
Note that even if we consider the thermal relaxation time in our model as, $\tau_\mathrm{diff} = 8~\mathrm{\mu s}$ (obtained from Fig.~\ref{fig:abs3rec1}), this duration time does not change greatly and is about $1.35~\mathrm{\mu s}$.

To achieve Kerr nonlinearity we must complete our operation before this time, namely the time that the light is absorbed by the material.  
This simple picture is straightforward to understand.  
If we are to obtain a faster operation speed we require a smaller $\tau_\mathrm{tot}$ because it determines the rise and fall time of the light energy in the cavity.  
$\tau_\mathrm{tot}$ is given by $\tau_{\mathrm{tot}}^{-1}=\tau_{\mathrm{loss}}^{-1}+\tau_{\mathrm{coup}}^{-1}+\tau_{\mathrm{abs}}^{-1}$, where $\tau_\mathrm{loss}$  and $\tau_\mathrm{coup}$ are the photon lifetimes defined by the loss rate that do not contribute to the generation of heat and coupling to the waveguides.  
Since $\tau_{\mathrm{abs}}$ and $\tau_{\mathrm{loss}}$ are mainly determined by the material and the structure that we use, the only parameter that we can control is $\tau_{\mathrm{coup}}$.  
This discussion suggests that optical Kerr operation is possible by controlling the coupling between the cavity and the waveguides, because it enables us to release light into the waveguides before it is absorbed by the material.

In the following section, we perform a rigorous analysis showing that Kerr operation is indeed possible by changing $\tau_{\mathrm{coup}}$.

\section{Rigorous modeling of the optical Kerr effect and thermo-optic effect in a toroid microcavity}\label{sec2}
\subsection{CMT in whispering gallery mode resonator}
\label{sec:CMT}
\begin{figure}[htbp]
\begin{center}
	\includegraphics*[width=2.8in]{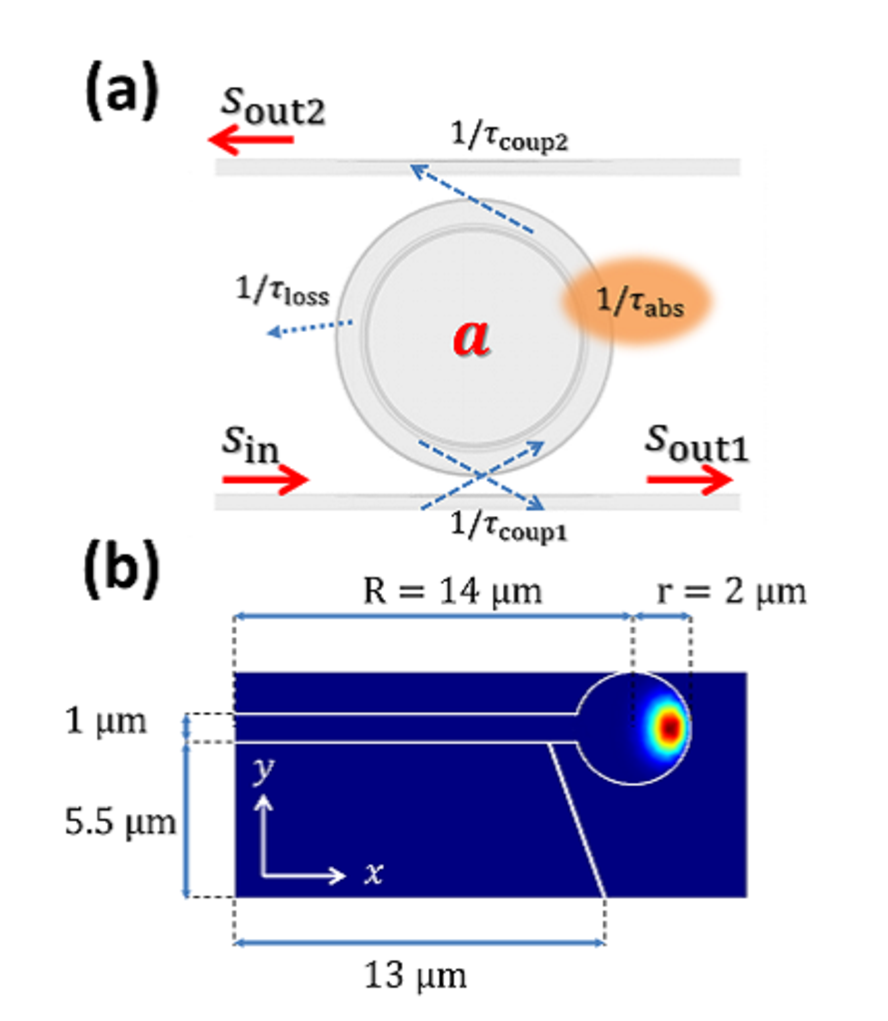}
	\caption{(Color online) (a) Cavity structure used for numerical analysis. (b) Cross-section of the mode intensity profile of a toroid microcavity. \cite{oxborrow2007a}}
	\label{fig:CMTbase1}
\end{center}
\end{figure}

First we describe our master equation based on the coupled mode theory in a whispering gallery mode (WGM) resonator to obtain the linear and nonlinear transmittance.  
The structure is shown in Fig.~\ref{fig:CMTbase1}(a).  
Because a two-port system (a side coupled cavity with one waveguide) makes the bistable operation difficult to observe, we focus on a side-coupled four-port system \cite{rokhsari2004b} throughout this paper.  
It consists of a toroid (ring) cavity and two waveguides for input and output light.  
a detailed discussion on the comparison between side-coupled two- and four- port systems will be provided elsewhere\cite{yoshiki2012a}.  
Briefly, as discussed in \S\ref{sec1} we need to make the coupling large in order to prevent heat accumulating in the system.  
However, the transmittance spectrum of a two-port system is shallower in an over-coupled configuration, which makes the switching contrast of the output light very low and difficult to distinguish.  
Hence, optical-bistable switching with high contrast is difficult to achieve in a two-port system in the presence of the TO effect.

The mode amplitude $a$ in the cavity is given as \cite{manolatou1999a},
\begin{eqnarray}
\frac{\mathrm{d}a}{\mathrm{d}t} &=& \left[ j \omega_0 - \frac{1}{2} \left( \frac{1}{\tau_{\mathrm{abs}}} + \frac{1}{\tau_{\mathrm{loss}}} + \frac{1}{\tau_{\mathrm{coup1}}} + \frac{1}{\tau_{\mathrm{coup2}}} \right) \right] a \nonumber\\
 && + \sqrt{\frac{1}{\tau_\mathrm{coup1}}}\exp{(j\theta)} s_\mathrm{in},
\label{eq:CMT23}
\end{eqnarray}
where $\omega_0$ is the resonant frequency of the cavity.  
The input wave $s_{\mathrm{in}}$ excites the counter clock-wise (CCW) mode in the cavity.  
We assume an ideal cavity where there is no coupling between the CW and CCW modes.  
$\tau_\mathrm{coup1}$ and $\tau_\mathrm{coup2}$ are photon lifetimes determined by the coupling with the lower and upper waveguides, respectively.  
$\theta$ is the relative phase between the mode amplitude in the cavity and the optical wave in the lower waveguide and is given as,
\begin{eqnarray}
\theta  = 4\pi^2 n_0 (R+r)\left( \frac{1}{\lambda_0} - \frac{1}{\lambda} \right).
\label{eq:theta1}
\end{eqnarray}
Here $R$, $r$, $\lambda$ and $\lambda_0$ are the major and minor radiuses of the cavity (shown in Fig.~\ref{fig:CMTbase1}(b)), the input wavelength and the resonant wavelength of the cavity, respectively.  
Equation~(\ref{eq:theta1}) shows that the phase between $a$ and $s_\mathrm{in}$ becomes unmatched on off-resonance.  
Output waves $s_\mathrm{out1}$ and $s_\mathrm{out2}$ are given as,
\begin{eqnarray}
s_{\mathrm{out1}} &=& \exp{\left( {-j\beta_1 d} \right)} \nonumber \\
&& \times \left[s_{\mathrm{in}} - \sqrt{\frac{1}{\tau_\mathrm{coup1}}}\exp{(-j\theta_1)} a\right] \\
s_{\mathrm{out2}} &=& \exp{\left( {-j\beta_2 d} \right)} \sqrt{\frac{1}{\tau_\mathrm{coup2}}} a,
\label{eq:CMT25}
\end{eqnarray}
where $\beta_1$ and $\beta_2$ are the propagation constants of the lower and upper waveguides and $d$ is the waveguide length.  
Note that the relative phase between the mode amplitude in the cavity $a$ and the upper waveguide $s_{\mathrm{in}}$ is always zero because there is no incident wave in the upper waveguide.  

When we use a slowly varying envelope approximation, i.e.,
\begin{eqnarray}
a(t) &=& A(t) \exp{\left( j \omega t \right)} \\
s_\mathrm{in}(t) &=& S_{\mathrm{in}}(t) \exp{\left( j \omega t \right)},
\label{eq:ap1}
\end{eqnarray}
we can rewrite Eq.~(\ref{eq:CMT23}) as,
\begin{eqnarray}
\frac{\mathrm{d}A(t)}{\mathrm{d}t} &=& \left[ j \frac{2\pi c}{n_0}\left(\frac{1}{\lambda_0} - \frac{1}{\lambda}\right)  \nonumber \right. \\
 && - \left. \frac{1}{2} \left( \frac{1}{\tau_{\mathrm{abs}}} + \frac{1}{\tau_{\mathrm{loss}}} + \frac{1}{\tau_{\mathrm{coup1}}} + \frac{1}{\tau_{\mathrm{coup2}}} \right) \right] A(t) \nonumber \\
 && + \sqrt{ \frac{1}{\tau_{\mathrm{coup1}}}} \exp{\left( j \theta \right)}S_{\mathrm{in}}(t).
\label{eq:master1}
\end{eqnarray}
where $A(t)$, $S_{\mathrm{in}}(t)$ and $\omega= 2 \pi c/(n_0 \lambda)$ are the envelopes of the cavity mode and the waveguide mode and the frequency of the input wave, respectively.  
Equation~(\ref{eq:master1}) is the master equation of the linear system.  
By using this equation, we now can calculate the energy in the cavity and the output power at an arbitrary time.  

\subsection{Modeling the nonlinearities}
To describe the nonlinear effects in our model, we take account of the nonlinear refractive index modulation caused by the Kerr effect ($\Delta n_{\mathrm{Kerr}}$) and TO effect ($\Delta n_\mathrm{TO}$) in the master equation (Note that the carrier-plasma effect is negligible in silica due to its large bandgap).  
The nonlinearities in an optical cavity result in a shift in the resonant wavelength because the optical path length changes.  
Thus, the shift of the resonant wavelength of a cavity is given as, 
\begin{eqnarray}
\delta \lambda(t) = \frac{\Delta n(t)}{n_0} \lambda_0,
\label{eq:l1}
\end{eqnarray}
where $\Delta n(t)$ is the effective nonlinear refractive index change of the cavity.  
By substituting Eq.~(\ref{eq:l1}) into Eqs.~(\ref{eq:master1}) and (\ref{eq:theta1}), we obtain
\begin{eqnarray}
\frac{\mathrm{d}A(t)}{\mathrm{d}t} &=& \left[ j \frac{2\pi c}{n_0 + \Delta n(t)}\left(\frac{1}{\lambda_0 + \delta \lambda (t)} - \frac{1}{\lambda}\right) \right. \nonumber \\
 && - \left. \frac{1}{2}\left( \frac{1}{\tau_{\mathrm{abs}}} + \frac{1}{\tau_{\mathrm{loss}}} + \frac{1}{\tau_{\mathrm{coup1}}} + \frac{1}{\tau_{\mathrm{coup2}}} \right) \right] A(t) \nonumber \\ 
&& + \sqrt{ \frac{1}{\tau_{\mathrm{coup1}}}} \exp{\left( j \theta \right)}S_{\mathrm{in}}(t).
\label{eq:master3}
\end{eqnarray}
\begin{eqnarray}
\theta  = 4\pi^2 (n_0 + \Delta n(t)) (R+r)\left( \frac{1}{\lambda_0 + \delta \lambda (t)} - \frac{1}{\lambda} \right).
\label{eq:theta2}
\end{eqnarray}
These are the master equations that we used in our model, which take the nonlinearities into account.

Next, we describe how we calculated the nonlinear refractive index change $\Delta n_{\mathrm{Kerr}}$ and $\Delta n_{\mathrm{TO}}$.  
We can directly calculate $\Delta n_{\mathrm{Kerr}}$ from Eq.~(\ref{eq:kerr9}).  
Taking the spatial dependency into account, we obtain,
\begin{eqnarray}
\Delta n_{\mathrm{Kerr}}(x,y,t) &=& n_2 I(x,y,t) \nonumber \\
 &=&\frac{2 n_2 c}{n_0} \tilde{U}_{\mathrm{p}}(x,y,t),
\label{eq:index1}
\end{eqnarray}
where $\tilde{U}_{\mathrm{p}}(x,y,t)$ is the energy density distribution of the cavity mode in $x, y$ cross-sectional coordinates.  
It is given as,
\begin{eqnarray}
\tilde{U}_{\mathrm{p}}(x,y,t) = \frac{U_{\mathrm{p}}(t)}{2\pi R} \tilde{I}(x,y),
\label{eq:dis3}
\end{eqnarray}
where $U_\mathrm{p} = |A(t)|^2$ is the energy of the light stored in the cavity and $\tilde{I}(x,y)$ is the normalized cross-sectional power density distribution of the whispering gallery mode obtained by FEM (see Ref.~\cite{oxborrow2007a}).  
It is normalized as $\int\!\!\!\int\tilde{I}(x,y)\mathrm{d}x\mathrm{d}y = 1$, and the profile is shown in Fig.~\ref{fig:CMTbase1}(b).

The refractive index change $\Delta n_{\mathrm{TO}}$, which is induced by the TO effect, is described as,
\begin{eqnarray}
\Delta n_{\mathrm{TO}}(x,y,t) = n \xi [T(x,y,t) - 300~[\mathrm{K}]].
\label{eq:index5}
\end{eqnarray}
The cross-sectional temperature distribution is calculated by using 2D-FEM (COMSOL Multiphysics).  
By setting the heat source in the dielectric cavity as,
\begin{eqnarray}
 Q'(x,y,t) = \left\{
  \begin{array}{ll}
   \tau_{\mathrm{abs}}^{-1} \tilde{U}_{\mathrm{p}}(x,y,t), & \textrm{in cavity,}\\
   0, & \textrm{in air,}\\
  \end{array}
 \right.
\end{eqnarray}
where $\tau_{\mathrm{abs}}^{-1}$ is the thermal generation rate caused by the material absorption, we can obtain the temperature $T(x,y,t)$ at any time and at any position by performing an FEM calculation.

Finally, we obtain the effective nonlinear refractive index change $\Delta n(t)$ as,
\begin{eqnarray}
\Delta n(t) = \frac{\int\!\!\!\int\left[ \Delta n_{\mathrm{TO}}(x,y,t) + \Delta n_{\mathrm{Kerr}}(x,y,t) \right]\tilde{I}(x,y) \mathrm{d}x \mathrm{d}y}{\int\!\!\!\int\tilde{I}(x,y) \mathrm{d}x \mathrm{d}y}. \nonumber \\
\label{eq:overlap}
\end{eqnarray}

Now by solving Eq.~(\ref{eq:master3}) sequentially using Eqs.~(\ref{eq:l1})--(\ref{eq:overlap}), we can obtain the light energy in the cavity $U_{\mathrm{p}}(t)$ and the output powers $P_\mathrm{out1} = |S_\mathrm{out1}|^2$ and $P_\mathrm{out2} = |S_\mathrm{out2}|^2$.

\subsection{Determining the absorption and the photon lifetimes}
\label{sec:photon}
The total photon lifetime $\tau_\mathrm{tot}$ is defined as,
\begin{eqnarray}
\tau_{\mathrm{tot}} &=& \left( \tau_{\mathrm{mat}}^{-1} + \tau_{\mathrm{water}}^{-1} + \tau_{\mathrm{cont}}^{-1} + \tau_{\mathrm{rad}}^{-1} + \tau_{\mathrm{surf}}^{-1} \right.\nonumber \\
  &&+ \left. \tau_{\mathrm{coup1}}^{-1} + \tau_{\mathrm{coup2}}^{-1} \right)^{-1},
\end{eqnarray}
where $\tau_{\mathrm{mat}}$, $\tau_{\mathrm{water}}$, $\tau_{\mathrm{cont}}$, $\tau_{\mathrm{rad}}$, $\tau_{\mathrm{surf}}$, $\tau_{\mathrm{coup1}}$ and $\tau_{\mathrm{coup2}}$ are photon lifetimes determined by the absorption of the material, water absorption that is usually present on the surface of the cavity, absorption caused by surface contamination, radiation loss, scattering loss, coupling to the lower waveguide and coupling to the upper waveguide, respectively.  
Since this expression is very complicated, we define the photon lifetime that is related to the absorption as, $\tau_\mathrm{abs}^{-1} = \tau_\mathrm{mat}^{-1} + \tau_\mathrm{water}^{-1}+ \tau_\mathrm{cont}^{-1}$, losses to the outside of the cavity $\tau_\mathrm{loss}^{-1} = \tau_\mathrm{rad}^{-1} + \tau_\mathrm{surf}^{-1}$, and the couplings $\tau_\mathrm{coup1,2}$.  
By using these photon lifetimes, we can simplify our analysis, because $\tau_\mathrm{abs}$ contributes on the generation of heat, but the other factors ($\tau_\mathrm{loss}$ and $\tau_\mathrm{coup1,2}$) do not.

Here we describe the photon lifetimes that we used in our analysis.  
The material absorption of silica at telecom wavelength is usually very small ($\alpha = 0.2~\mathrm{dB/km}$ \cite{miya1979a}), but, it is known that silica toroid microcavities have much larger absorption due to the water layer and contamination on their surfaces.  
And the $Q$ factor is limited by these absorptions in practice \cite{collot1993a}.  
Thus, we decided to consider two cases, which we call an ideal case and a realistic (worst) case.  
In an ideal case, we assume that the absorption loss is determined by only the material (this means that the absorption loss is extremely small) and the experimental $Q$ is limited by the losses to the outside of the cavity $\tau_\mathrm{loss}^{-1}$.  
On the other hand, we assume that the experimental $Q$ is limited by the absorption loss $\tau_\mathrm{abs}^{-1}$ in a realistic case.  
This implies the worst case in terms of thermal accumulation, because a large part of the incident light eventually turns into heat.  
Any result should be better than that obtained with this latter condition.

To consider those two cases, first we fix the intrinsic photon lifetime $\tau_{\mathrm{int}} = (\tau_{\mathrm{abs}}^{-1} + \tau_{\mathrm{loss}}^{-1})^{-1}$ as $329~\mathrm{ns}$ (corresponding to $Q_\mathrm{int} = 4 \times 10^8$ \cite{toroid2}, where $Q_\mathrm{int}$ is the intrinsic $Q$), since this is the record largest experimental $\tau_\mathrm{int}$.
In the ideal case we disregard losses than the intrinsic material absorption; so we use $\tau_\mathrm{abs}=164~\mathrm{\mu s}$ (corresponding to $Q_\mathrm{abs} \simeq Q_\mathrm{mat} = 2 \times 10^{11}$ \cite{maleki2004a}) and $\tau_\mathrm{loss}=330~\mathrm{ns}$; i.e. $\tau_\mathrm{int}^{-1} \simeq \tau_\mathrm{loss}^{-1} \gg \tau_\mathrm{abs}^{-1}$.  
On the other hand, we use $\tau_\mathrm{abs}=329~\mathrm{ns}$ (corresponding to $Q_\mathrm{abs} = 4 \times 10^{8}$ \cite{rokhsari2004a, toroid2}) and $\tau_\mathrm{loss}=\infty$ for the realistic case.  
This $Q_\mathrm{abs}$ value is the same as the highest experimental $Q_\mathrm{int}$, for the reason discussed above; i.e. $\tau_\mathrm{int}^{-1} \simeq \tau_\mathrm{abs}^{-1} \gg \tau_\mathrm{loss}^{-1}$.

Although $\tau_\mathrm{int}$ is determined by the material and structure, we can control $\tau_{\mathrm{coup1}}$ and $\tau_{\mathrm{coup2}}$ by adjusting the distance between the fiber and the cavity \cite{spillane2003a}.  
With this in mind, we conducted numerical simulations for various $\tau_{\mathrm{coup2}}$ values.  
Note that $\tau_\mathrm{coup1}$ is controlled to satisfy $\tau_\mathrm{coup1} = (\tau_\mathrm{int}^{-1} + \tau_\mathrm{coup2}^{-1})^{-1}$ and thus achieve critical coupling \cite{manolatou1999a,cai2000a,yanik2003a} between the cavity and the lower waveguide.  
In this condition, the power transmittance through the lower waveguide $T_\mathrm{out1} = P_\mathrm{out1} / P_\mathrm{in}$ decreases to zero on resonance and high contrast can be obtained between two output states\cite{yoshiki2012a}.

\begin{figure*}[htbp]
 \begin{center}
 \includegraphics*[width=6in]{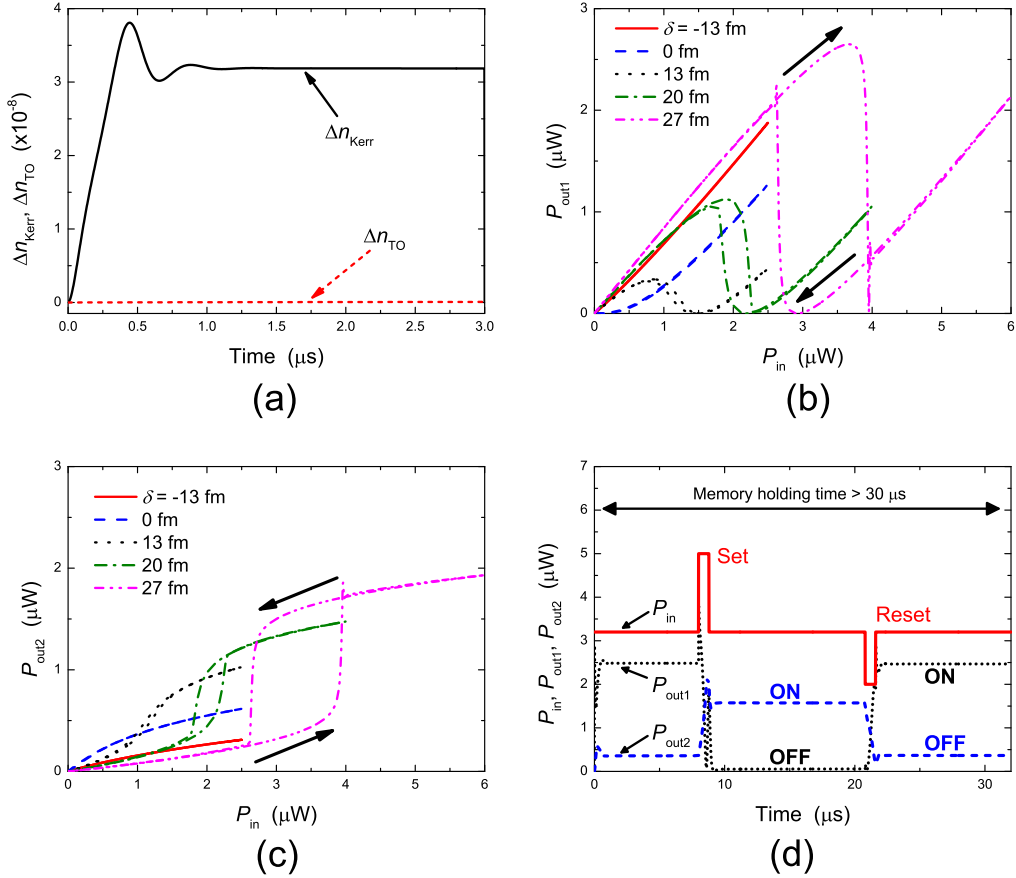}
 \caption{(Color online) (a) $\Delta n_{\mathrm{Kerr}}$ and $\Delta n_{\mathrm{TO}}$ vs. time when a $6~\mathrm{\mu W}$ rectangular pulse is inputted.  
The wavelength detuning $\delta$ between the input light and the initial cavity resonance is $27~\mathrm{fm}$.  
$\tau_{\mathrm{coup2}}$ is set equal to $\tau_\mathrm{int}$.  
(b) $P_{\mathrm{out1}}$ vs. $P_{\mathrm{in}}$ of a triangular input pulse with different $\delta$ values, where $P_{\mathrm{out1}}=|s_\mathrm{out1}|^2$.  
(c) $P_{\mathrm{out2}}$ vs. $P_{\mathrm{in}}$ with different $\delta$, where $P_{\mathrm{out2}}=|s_\mathrm{out2}|^2$.  
$\delta = 13$, 20, and 27~fm correspond to $(\sqrt{3}/2) \Delta \lambda_{\mathrm{FWHM}}$, $(3 \sqrt{3}/4) \Delta \lambda_{\mathrm{FWHM}}$, and $\sqrt{3} \Delta \lambda_{\mathrm{FWHM}}$, respectively.  
$\Delta \lambda_{\mathrm{FWHM}}$ is the full-width at half-maximum of the cavity resonant spectrum width. ($\Delta \lambda_{\mathrm{FWHM}} \simeq 15$~fm).  
(d) Optical bistable memory operation.  The solid, dotted and broken lines represent $P_{\mathrm{in}}$, $P_{\mathrm{out1}}$ and $P_{\mathrm{out2}}$, respectively.  }
  \label{fig:abs2rec1}
 \end{center}
\end{figure*}
\section{Numerical calculations}
\subsection{An ideal case: Small material absorption}
\label{sec:ideal}
In this section, we show that Kerr bistable memory is easily feasible, without careful adjustment of $\tau_{\mathrm{coup}}$, if only the inherent material absorption of silica is present.

First, we input a rectangular pulse to investigate the refractive index changes $\Delta n_{\mathrm{Kerr}}$ and $\Delta n_{\mathrm{TO}}$.  
The result is shown in Fig.~\ref{fig:abs2rec1}(a), where $\tau_{\mathrm{coup2}}$ is equal to $\tau_\mathrm{int}$.  
It shows that $\Delta n_{\mathrm{Kerr}}$ is always larger than $\Delta n_{\mathrm{TO}}$, which tells us that the influence of the absorption induced thermal generation is nearly negligible.  
Hence the Kerr effect is easily obtained without it suffering from the TO effect.

Next, we input a triangular pulse to investigate the relationship between the input and output of the system.  
To allow us to charge and discharge the cavity gradually, we set the pulse rising/falling rate of the triangular inputs at  $\mathrm{d}P_\mathrm{in}/\mathrm{d}t=62.5~\mathrm{nW/\mu s}$.  
Figure~\ref{fig:abs2rec1}(b) and (c) are plotted from the input-output response of a triangular input of different detuning values $\delta$.  
The lower coupling photon lifetime $\tau_{\mathrm{coup1}}$ is set equal to $(\tau_\mathrm{int}^{-1} + \tau_\mathrm{coup2}^{-1})^{-1}$ to achieve critical coupling.  When $\delta$ is greater than 20~fm, clear hysteresis is observed, which is direct evidence of optical bistability.  
Optical bistability is observed in the longer side of the wavelength detuning, because the Kerr effect increases the refractive index (shifts the resonance toward the longer wavelength).  
Note that we obtained a large contrast between the two bistable states in Fig.~\ref{fig:abs2rec1}(b) because the cavity transmittance $P_\mathrm{out1}$ falls to zero on resonance in the critical coupling \cite{yanik2003a}.

Finally, we performed optical memory operations as shown in Fig.~\ref{fig:abs2rec1}(d).  
Again, we set $\tau_{\mathrm{coup2}}$ equal to $\tau_\mathrm{int}$ and the detuning $\delta$ equal to $27~\mathrm{fm}$.  
The solid line is an input with a drive power $P_{\mathrm{in}}^{\mathrm{drive}}$ of $3.2 \ \mathrm{\mu W}$.  
The peak power of the $0.8$-$\mathrm{\mu s}$ square set pulse is $P_{\mathrm{in}}^{\mathrm{set}}=5\ \mathrm{\mu W}$.  
To reset the system, we reduce the input power to $P_{\mathrm{in}}^{\mathrm{reset}}=2\ \mathrm{\mu W}$ for a duration of $0.8$~$\mathrm{\mu s}$, which we call a reset pulse.  
The duration of the negative reset pulse must be longer than the discharging time of the cavity, which is equal to $\tau_{\mathrm{tot}}$; otherwise the cavity does not reset.  
Set and reset pulses are inputted at $t=8$ and $20.8~\mathrm{\mu s}$.   $P_{\mathrm{out2}}$ (shown as the broken line) rises to high (ON) state when the set pulse is inputted.  
It keeps the ON state until the reset pulse is injected.  
After the reset pulse has been entered, $P_{\mathrm{out2}}$ drops to low (OFF) state and holds this state.  
$P_\mathrm{out1}$ (indicated by the dotted line in Fig.~\ref{fig:abs2rec1}(d)) shows the inverse behavior of $P_\mathrm{out2}$.  
Figure~\ref{fig:abs2rec1}(d) clearly shows optical memory operation, which is based on Kerr nonlinearity.  
Although the ``memory holding time'' demonstrated with this calculation (shown in Fig.~\ref{fig:abs2rec1}(d)) is about $30~\mathrm{\mu s}$, it can be much larger, since $P_{\mathrm{out1}}$ and $P_{\mathrm{out2}}$ exhibit almost an plateau response due to the small material absorption.

\subsection{A realistic case:  Large material absorption}
\label{sec:real}
As shown in \S\ref{sec:ideal}, the realization of a Kerr bistable memory is feasible without it suffering from the TO effect, even when the coupling is not large, if only the inherent absorption of silica is present in the cavity.  
In reality, however, other sources of absorption occur in a microcavity, such as surface absorptions caused by water and contamination.  
Thus, in this section, we use $\tau_{\mathrm{abs}} = 329~\mathrm{ns}$ to simulate a case with faster thermal generation.  
As discussed above, here we assume that the $Q_{\mathrm{int}}$ of the cavity is limited by absorption and not by losses to the outside of the cavity, since it appears to be the worst (but realistic case) for toroid microcavities.  
We employ $\tau_{\mathrm{abs}}$, which is derived from the highest experimental $Q_{\mathrm{int}}$ ($\tau_{\mathrm{abs}} = 329~\mathrm{ns}$ corresponds to $Q_{\mathrm{int}}=4\times10^8$).  
If we can clarify the requirements for demonstrating a Kerr bistable memory under this condition, it is a significant step toward the experimental realization of a Kerr bistable memory in silica toroid microcavities.

\begin{figure}[htbp]
 \begin{center}
  \includegraphics*[width=2.8in]{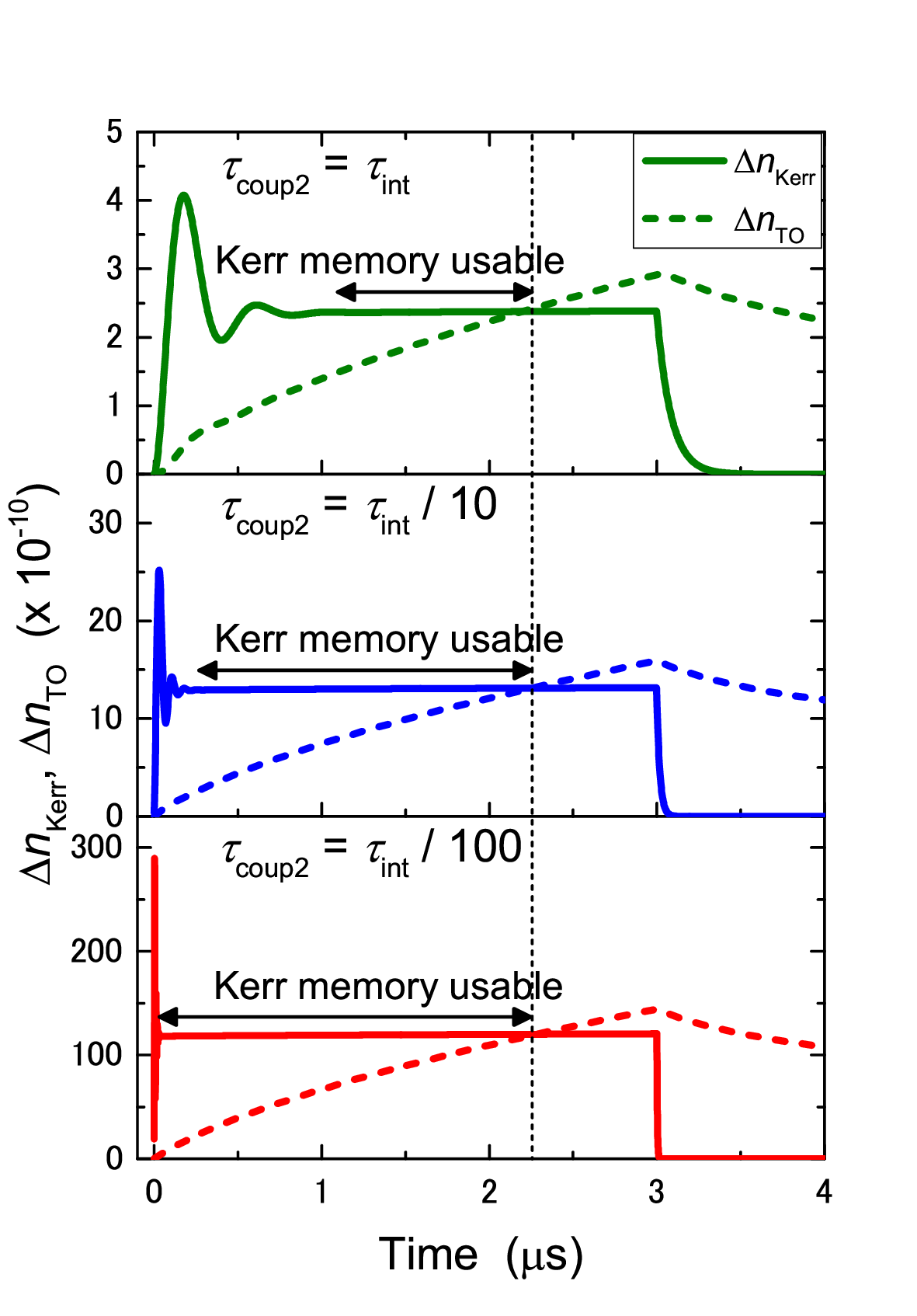}
  \caption{(Color online) $\Delta n_{\mathrm{Kerr}}$(solid line) and $\Delta n_{\mathrm{TO}}$(broken line) vs. time for different $\tau_{\mathrm{coup2}}$ values.  
The input pulse power is set to $500 \times (Q_\mathrm{tot}^{\tau_\mathrm{coup2}= \tau_\mathrm{int}/100}/Q_\mathrm{tot})^2$ in $\mathrm{\mu W}$.  
The detuning $\delta$ is set at $\sqrt{3} \Delta \lambda_{\mathrm{FWHM}}$, where $\Delta \lambda_{\mathrm{FWHM}}$ is 15~fm, 85~fm, and 782~pm when $\tau_{\mathrm{coup2}}$ is $\tau_\mathrm{int}$, $\tau_\mathrm{int}/10$, and $\tau_\mathrm{int}/100$, respectively.}
  \label{fig:abs3rec1}
 \end{center}
\end{figure}

First, in a similar way to that shown in Fig.~\ref{fig:abs2rec1}(a), we employ a rectangular pulse to obtain the refractive index change $\Delta n_{\mathrm{Kerr}}$ and $\Delta n_{\mathrm{TO}}$.  
The calculation results are shown in Fig.~\ref{fig:abs3rec1} for three different $\tau_{\mathrm{coup2}}$ values ($\tau_\mathrm{coup1}$ is adjusted to satisfy $\tau_\mathrm{coup1}^{-1} = \tau_\mathrm{coup2}^{-1} + \tau_{\mathrm{int}}^{-1}$).
Figure~\ref{fig:abs3rec1} shows that $\Delta n_{\mathrm{TO}}$ is larger than $\Delta n_{\mathrm{Kerr}}$ in all three cases when $t$ is larger than $\sim 2.3~\mathrm{\mu s}$.  This number gives us the upper limit of the Kerr memory holding time without the memory suffering from the TO effect.  It also tells us that this number is insensitive to $\tau_{\mathrm{coup2}}$.  This result is consistent with Eq.~(\ref{eq:eq12}) obtained from a simple model, where the equation is dependent on $\tau_{\mathrm{abs}}$ but independent of $\tau_\mathrm{coup}$.@ Figure~\ref{fig:abs3rec1} also shows the effect of the different charging speed resulted by different $\tau_{\mathrm{coup2}}$ values.  The cavity charging time is much faster for $\tau_{\mathrm{coup2}}=\tau_\mathrm{int} / 100$, which allows the cavity to reach a plateau $\Delta n_{\mathrm{Kerr}}$ domain much faster.  This enables us to have a longer ``Kerr memory usable'' regime, which allows us to use the cavity for longer as an optical Kerr memory.  Figure~\ref{fig:abs3rec1} shows that we can maximize the Kerr dominant ``Kerr memory usable'' regime by setting $\tau_{\mathrm{coup2}}$ as small as possible.  Again, this result is in consistent with that obtained from the simple analysis described in \S~\ref{sec1}.  

Next, we show on how we can obtain Kerr optical bistability in a large absorption condition.
\begin{figure}[htbp]
 \begin{center}
  \includegraphics*[width=2.8in]{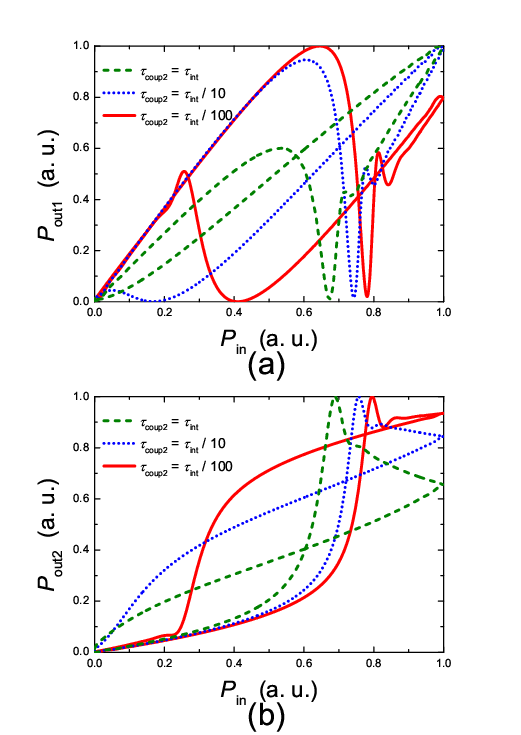}
  \caption{(Color online) Input-output response of a triangular pulse input.  
(a) $P_{\mathrm{out1}}$ vs. $P_{\mathrm{in}}$ for different $\tau_{\mathrm{coup2}}$.  
(b) $P_{\mathrm{out2}}$ vs. $P_{\mathrm{in}}$ for different $\tau_{\mathrm{coup2}}$.  
The $x$-axis and $y$-axis are normalized as $P_{\mathrm{out1}}^{\mathrm{max}}=P_{\mathrm{out2}}^{\mathrm{max}}=P_{\mathrm{in}}^{\mathrm{max}}=1$.  
The detuning $\delta$ is set at $\sqrt{3} \Delta \lambda_{\mathrm{FWHM}}$.  
The input pulse width is $15~\mathrm{\mu s}$, $3~\mathrm{\mu s}$ and $0.3~\mathrm{\mu s}$ and the pulse height is $4.5~\mathrm{\mu W}$, $160~\mathrm{\mu W}$, $15~\mathrm{m W}$ when $\tau_{\mathrm{coup2}}$ is $\tau_\mathrm{int}$, $\tau_\mathrm{int}/10$, and $\tau_\mathrm{int}/100$, respectively. }
  \label{fig:4}
 \end{center}
\end{figure}
Figure~\ref{fig:4}(a) and (b) show the input and output power relationships ($P_{\mathrm{out1}}$ and $P_{\mathrm{out2}}$) for different $\tau_{\mathrm{coup2}}$ values when a triangular pulse is inputted.  
As shown in Fig.~\ref{fig:4}(a) and (b), a hysteresis loop is observed when $\tau_{\mathrm{coup2}}=\tau_\mathrm{int}/100$, but the loops are deformed when $\tau_{\mathrm{coup2}} \geq \tau_\mathrm{int}/10$.  
This is because the light cannot charge and discharge the cavity quickly enough before the heat accumulates in the system when $\tau_{\mathrm{coup2}}$ is large.   Again, we obtained a clear hysteresis loop only when we made the coupling strong (i.e. $\tau_{\mathrm{coup2}}$ is small).

\begin{figure}[htbp]
 \begin{center}
  \includegraphics*[width=2.8in]{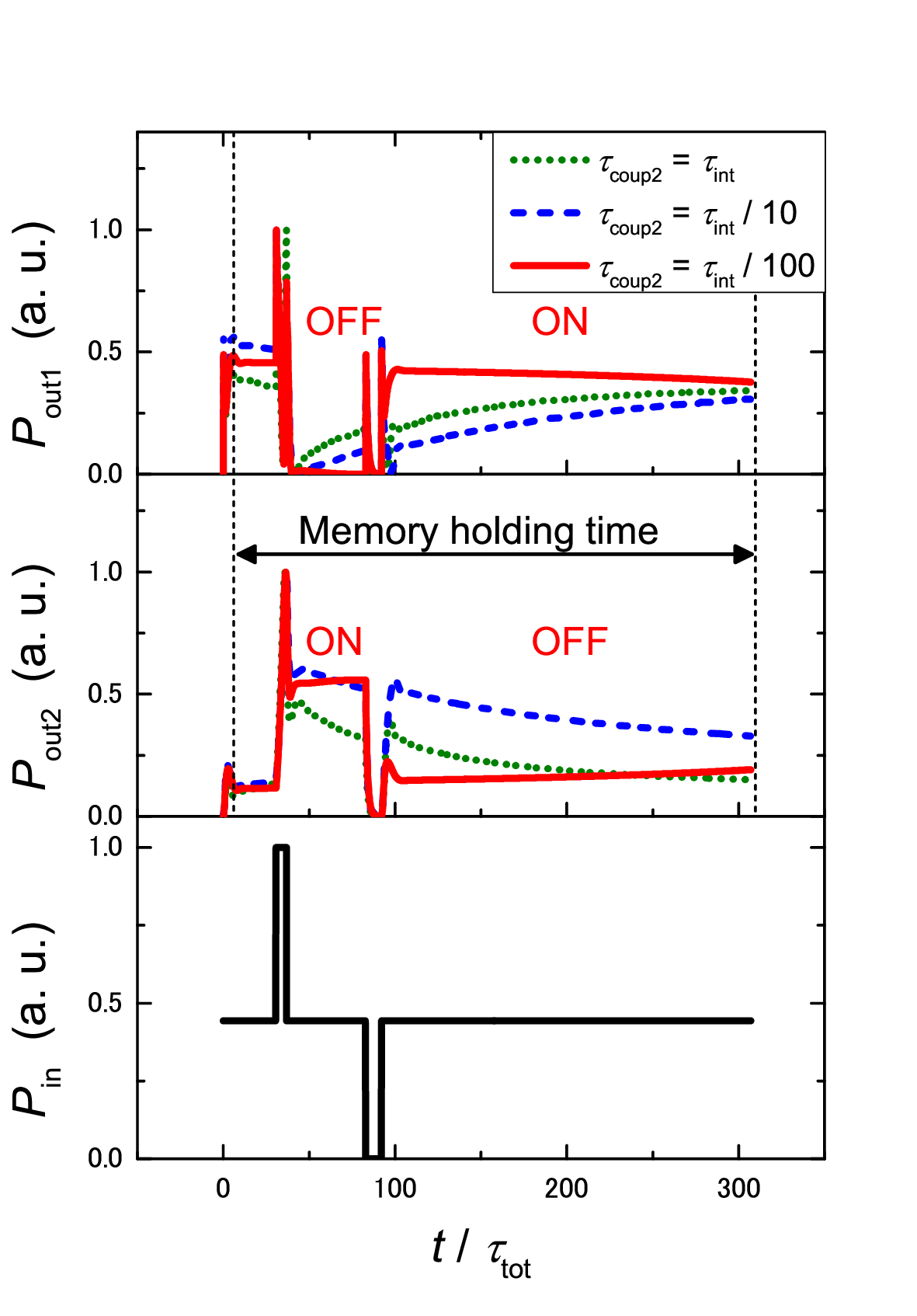}
  \caption{(Color online) Optical memory operation, when $\delta$ is $\sqrt{3}\Delta_\mathrm{FWHM}$.  
The input power of the drive light is 2 $\mathrm{\mu}$W, 80 $\mathrm{\mu}$W, and 7.3 mW,
when $\tau_\mathrm{coup2}$ is $\tau_\mathrm{int}$, $\tau_\mathrm{int}/10$ and $\tau_\mathrm{int}/100$, respectively.  
The horizontal axis is normalized with the photon lifetime $\tau_{\mathrm{tot}}$.  }
  \label{fig:5}
 \end{center}
\end{figure}

Finally, the memory operation for various $\tau_{\mathrm{coup2}}$ values is shown in Fig.~\ref{fig:5}.  
Since the response speed of $P_{\mathrm{out1}}$ and $P_{\mathrm{out2}}$ depend on the total photon lifetime $\tau_{\mathrm{tot}}$, we normalized the temporal axis $t$ by $\tau_{\mathrm{tot}}$.  
Figure~\ref{fig:5} shows clearly that the reset pulse does not work when $\tau_{\mathrm{coup2}} = \tau_\mathrm{int}$ and $\tau_\mathrm{int}/10$, and the memory operation cannot be obtained under this condition.  
If the system is operating in the Kerr dominant regime,  we should be able to reset the state by injecting a negative reset pulse.  
The pulse width needed for the reset pulse is $>\tau_\mathrm{tot}$, since we can discharge the cavity within this time.  
However, Fig.~\ref{fig:5} shows that significant heat is accumulating in the system, which prevents the system from resetting because $\Delta n_\mathrm{TO}$ cannot be reset by such a short negative pulse due to its much longer relaxation time.

On the other hand, when $\tau_{\mathrm{coup2}} \le \tau_\mathrm{int}/100$, we can successfully set and reset the system, and use the device as a Kerr bistable memory.  
However, the TO effect cannot be eliminated completely even in this case, and thus a holding time exists.  
$P_\mathrm{out1}$ is automatically switched from ON to OFF, or vice versa for $P_\mathrm{out2}$, due to the thermal accumulation.  
Figure ~\ref{fig:5} shows that the memory holding time for the realistic case is about 500~ns.  This value is inconsistent to the length of ``Kerr memory usable'' regime shown in Fig.~\ref{fig:abs3rec1} since the thermal nonlinearity depends on the accumulation of $U_\mathrm{abs}$ while the Kerr nonlinearity depend on the instantaneous $U_\mathrm{p}$.  
 
In this section we achieved Kerr bistable memory operation and obtained a sufficiently long memory holding time of 500~ns by allowing the system to charge and discharge quickly by adjusting $\tau_{\mathrm{coup2}}$.

\section{Discussion: Power consumption}
\label{sec:discussion}
We showed in previous sections that Kerr bistable operation is possible by adopting a large coupling constant.  
The basic idea is to allow the light to charge and discharge before it turns into heat.  
However, stronger coupling with waveguides (i.e. smaller $\tau_{\mathrm{coup2}}$) results in a lower total $Q$, which decreases the photon density in a cavity and makes the nonlinearity small.  
Hence, there is a trade-off between operating speed and operating power.  
\begin{table*}[htbp]
 {\bf \caption{Comparison of the power consumption of different systems}\label{table:1}}
 \begin{center}
  {\scriptsize
  \begin{tabular}{llllllll}
   \hline
   System type & Material & Nonlinearity & $Q_{\mathrm{int}}$ & $P_{\mathrm{drive}}$ & $U_{\mathrm{cons}} ^{(a)}$ & Exp./Cal. & Refs. \\ \hline
   Microring cavity& Si &  Thermo-optic  & $1.43\times10^5$ & $800~\mathrm{\mu W}$ & $3.7\times 10^{-12}~\mathrm{J}$ & Exp.& Ref.~\cite{almeida2004b} \\
   Photonic crystal & InGaAsP & Carrier-plasma  & $1.3\times10^5$ & $250~\mathrm{\mu W}$ & - & Exp. & Ref.~\cite{shinya2008a} \\
   Photonic crystal & Semiconductor & Kerr  & 557 & 133~mW & $2.0\times 10^{-17(b)}\mathrm{J}$ & Calc. & Ref.~\cite{soljacic2002a} \\
   Toroid microcavity & $\mathrm{SiO_2}$ & Kerr & $4\times10^8$ & 7.3~mW & $2.6\times 10^{-23(b)}\mathrm{J}$  & Calc. & This work. \\
   \hline
  \end{tabular}
  }
 \end{center}
 \footnotesize $^a$ Estimated from the cavity resonance shift of $(1/2)\Delta \lambda_\mathrm{FWHM}$.  We assume no TO and no carrier diffusion. \\
 \footnotesize $^b$ The energy change caused by the resonant wavelength shift $\Delta U = U_\mathrm{HWHM} ((1/2)\Delta \lambda_\mathrm{FWHM} / (\lambda_0 + (1/2)\Delta \lambda_\mathrm{FWHM}))$ is regarded as $U_\mathrm{cons}$, where $U_\mathrm{HWHM}$ is the energy that can cause a resonant wavelength shift of $(1/2)\Delta \lambda_\mathrm{FWHM}$ by using the Kerr effect. 
\end{table*}  

Now we consider two measures for evaluating the loss of our system; $Q_{\mathrm{int}}$ and $U_{\mathrm{cons}}$.  $Q_{\mathrm{int}}$ is the figure of merit of a cavity, whose value gives the cavity loss, and $U_{\mathrm{cons}}$ is the energy consumed for the operation.  We often want to increase $Q_{\mathrm{int}}$ and reduce $U_{\mathrm{cons}}$ to build a lossless system.  
Table~\ref{table:1} compares these values for different systems.  A bistable memory based on an ultrahigh-$Q$ silica toroid microcavity with Kerr nonlinearity has a clear advantage over other schemes in terms of both losses.  This lossless nature of this memory is the advantage of this system.

First, a system with a high $Q_\mathrm{int}$ yields a low loss, because $Q_\mathrm{int}$ corresponds to the fundamental loss characteristics of a cavity.  As shown in Table~\ref{table:1}, the $Q_\mathrm{int}$ of a toroid cavity is much higher than that of other types of cavities.  Furthermore, we would like to note that both linear loss and nonlinear losses are significant in other systems, especially those systems that use TO and carrier-based nonlinearities to achieve bistability.  This is because carrier generation is unavoidable.  Free carrier absorption (FCA) significantly decreases both the $Q_\mathrm{int}$ and the transmittance of the system \cite{barclay2005a}.  On the other hand, a Kerr bistable memory does not generate carriers and hence the system does not suffer from FCA loss.

Secondly, $U_{\mathrm{cons}}$ is extremely low in our system because Kerr nonlinearity does not absorb photons.  $U_{\mathrm{cons}}$ is the energy consumed for a bistable operation.  We can estimate $U_{\mathrm{cons}}$ from the required refractive index change, which is needed to obtain the cavity resonance shift for the operation.  The values are large in memories based on TO and carrier-plasma effects even though they have large nonlinear efficiencies, which is because they incorporate photon absorption.  On the other hand, the total photon number for the Kerr effect remains unchanged after the operation.  Only the wavelength (energy) of the photons changes, which results in an extremely low energy consumption.  Hence $U_{\mathrm{cons}}$ is almost negligible as shown in Table~\ref{table:1}.

As shown in Fig.~5, we need $P_{\mathrm{in}}^{\mathrm{drive}}=7.3$~mW to drive the system, which is significantly smaller than the value shown in Ref.~\cite{soljacic2002a}, which uses the same Kerr nonlinearity.  
This is due to the high $Q_\mathrm{int}$ of our system.  
However, this value is still larger than the experimentally demonstrated value in a photonic crystal nanocavity using carrier-based nonlinearity\cite{nozaki2012a}.  
Since the coefficients of carrier nonlinearities in semiconductor materials are normally larger than the Kerr nonlinearity in $\mathrm{SiO_2}$, it is not an easy task to reduce the driving power of our system to the same level.  
However, glass performs low propagation loss \cite{bauters2011a, lee2012a}, which is usually difficult to obtain with devices made of semiconductors.  

Although the driving power $P_{\mathrm{drive}}$ of our system is not necessarily the smallest, the linear, nonlinear and consumption losses are significantly smaller than those of other devices.  
With this in mind, this device is attractive for applications that require high efficiency $\eta = P_{\mathrm{out}} / P_{\mathrm{in}}$, such as quantum information processing\cite{politi2008a}.

\section{Conclusion}
We rigorously modeled the Kerr and the TO effects in a silica toroid microcavity by combining CMT and FEM.  We gained a clear understanding of the impact of adjusting the coupling, and showed that Kerr optical bistable memory operation is possible by adjusting the coupling between the cavity and the waveguides.  The memory holding time was about 500~ns.  Although the driving power was 7.3~mW, the energy consumed by the the system was extremely low.  This is because unlike other nonlinearities such as carrier of TO effects Kerr nonlinearity does not absorb photons.  In addition due to the ultrahigh-$Q$ of the system, the energy loss outside the system is also low.  Our Kerr bistable memory in a silica toroid microcavity exhibits extremely low loss and thus is suitable for applications such as quantum signal processing.
\section*{Acknowledgment}
Part of this work was supported by the Strategic Information and Communications R\&D Promotion Programme (SCOPE), the Canon Foundation, and the Support Center for Advanced Telecommunications Technology Research, Foundation.

\bibliographystyle{osajnl}{

\begin{thebibliography}{10}
\newcommand{\enquote}[1]{``#1''}

\bibitem{tsuda1990a}
H.~Tsuda and T.~Kurokawa, \enquote{Construction of an all-optical flip-flop by
  combination of 2 optical triodes,} Appl. Phys. Lett. \textbf{{57}},
  {1724--1726} ({1990}).

\bibitem{gibbs1985a}
H.~Gibbs, \emph{Optical bistability: controlling light with light} (Academic
  Press, 1985).

\bibitem{noda2003a}
Y.~Akahane, T.~Asano, B.~Song, and S.~Noda, \enquote{{High-Q photonic
  nanocavity in a two-dimensional photonic crystal},} Nature \textbf{{425}},
  {944--947} ({2003}).

\bibitem{tanabe2007a}
T.~Tanabe, M.~Notomi, E.~Kuramochi, A.~Shinya, and H.~Taniyama,
  \enquote{{Trapping and delaying photons for one nanosecond in an ultrasmall
  high-Q photonic-crystal nanocavity},} Nat. Photonics \textbf{{1}}, {49--52}
  ({2007}).

\bibitem{almani2003a}
D.~Armani, T.~Kippenberg, S.~Spillane, and K.~Vahala, \enquote{{Ultra-high-Q
  toroid microcavity on a chip},} Nature \textbf{{421}}, {925--928} ({2003}).

\bibitem{vahala2003a}
K.~Vahala, \enquote{{Optical microcavities},} Nature \textbf{{424}}, {839--846}
  ({2003}).

\bibitem{soljacic2002a}
M.~Soljacic, M.~Ibanescu, S.~Johnson, Y.~Fink, and J.~Joannopoulos,
  \enquote{{Optimal bistable switching in nonlinear photonic crystals},} Phys.
  Rev. E \textbf{{66}} ({2002}).

\bibitem{collot1993a}
L.~Collot, V.~Lefevreseguin, M.~Brune, J.~Raimond, and S.~Haroche,
  \enquote{Very high-{Q} whispering-gallery mode resonances observed on
  fused-silica microspheres,} Europhys. Lett. \textbf{{23}}, {327--334}
  ({1993}).

\bibitem{notomi2005a}
M.~Notomi, A.~Shinya, S.~Mitsugi, G.~Kira, E.~Kuramochi, and T.~Tanabe,
  \enquote{{Optical bistable switching action of Si high-Q photonic-crystal
  nanocavities},} Opt. Express \textbf{{13}}, {2678--2687} ({2005}).

\bibitem{notomi2009a}
L.-D. Haret, T.~Tanabe, E.~Kuramochi, and M.~Notomi, \enquote{{Extremely low
  power optical bistability in silicon demonstrated using 1D photonic crystal
  nanocavity},} Opt. Express \textbf{{17}}, {21108--21117} ({2009}).

\bibitem{almeida2004b}
V.~Almeida and M.~Lipson, \enquote{{Optical bistability on a silicon chip},}
  Opt. Lett. \textbf{{29}}, {2387--2389} ({2004}).

\bibitem{tanabe2005a}
T.~Tanabe, M.~Notomi, S.~Mitsugi, A.~Shinya, and E.~Kuramochi,
  \enquote{{All-optical switches on a silicon chip realized using photonic
  crystal nanocavities},} Appl. Phys. Lett. \textbf{{87}}, 151112 ({2005}).

\bibitem{shinya2008a}
A.~Shinya, S.~Matsuo, Yosia, T.~Tanabe, E.~Kuramochi, T.~Sato, T.~Kakitsuka,
  and M.~Notomi, \enquote{{All-optical on-chip bit memory based on ultra high Q
  InGaAsP photonic crystal},} Opt. Express \textbf{16}, 19382--19387 (2008).

\bibitem{nozaki2012a}
K.~Nozaki, A.~Shinya, S.~Matsuo, Y.~Suzaki, T.~Segawa, T.~Sato, Y.~Kawaguchi,
  R.~Takahashi, and M.~Notomi, \enquote{{Ultralow-power all-optical RAM based
  on nanocavities},} Nat. Photonics \textbf{{6}}, {248--252} ({2012}).

\bibitem{spillane2005a}
M.~Spillane, T.~Kippenberg, K.~Vahala, K.~Goh, E.~Wilcut, and H.~Kimble,
  \enquote{{Ultrahigh-$Q$ toroidal microresonators for cavity quantum
  electrodynamics},} Phys. Rev. A \textbf{{71}}, {013817} ({2005}).

\bibitem{yoshie2004a}
T.~Yoshie, A.~Scherer, J.~Hendrickson, G.~Khitrova, H.~Gibbs, G.~Rupper,
  C.~Ell, O.~Shchekin, and D.~Deppe, \enquote{{Vacuum Rabi splitting with a
  single quantum dot in a photonic crystal nanocavity},} Nature \textbf{{432}},
  {200--203} ({2004}).

\bibitem{politi2008a}
A.~Politi, M.~J. Cryan, J.~G. Rarity, S.~Yu, and J.~L. O'Brien,
  \enquote{{Silica-on-silicon waveguide quantum circuits},} {Science}
  \textbf{{320}}, {646--649} ({2008}).

\bibitem{takesue2007a}
H.~Takesue, Y.~Tokura, H.~Fukuda, T.~Tsuchizawa, T.~Watanabe, K.~Yamada, and
  S.-I. Itabashi, \enquote{{Entanglement generation using silicon wire
  waveguide},} Appl. Phys. Lett. \textbf{{91}} ({2007}).

\bibitem{kawashima2008a}
H.~Kawashima, Y.~Tanaka, N.~Ikeda, Y.~Sugimoto, T.~Hasama, and H.~Ishikawa,
  \enquote{Optical bistable response in {AlGaAs}-based photonic crystal
  microcavities and related nonlinearities,} IEEE J. Quantum Electron.
  \textbf{{44}}, {841--849} ({2008}).

\bibitem{eggleton2011a}
B.~J. Eggleton, B.~Luther-Davies, and K.~Richardson, \enquote{{Chalcogenide
  photonics},} Nat. Photonics \textbf{{5}}, {141--148} ({2011}).

\bibitem{delong1989a}
K.~Delong, K.~Rochford, and G.~Stegeman, \enquote{Effect of 2-photon absorption
  on all-optical guided-wave devices,} Appl. Phys. Lett. \textbf{{55}},
  {1823--1825} ({1989}).

\bibitem{kang1994a}
J.~Kang, A.~Villeneuve, M.~Sheikbahae, G.~Stegeman, K.~Alhemyari, J.~Aitchison,
  and C.~Ironside, \enquote{Limitation due to 3-photon absorption on the useful
  spectral range for nonlinear optics in {AlGaAs} below half band-gap,} Appl.
  Phys. Lett. \textbf{{65}}, {147--149} ({1994}).

\bibitem{agrawal1995a}
G.~Agrawal, \emph{Nonlinear fibre optics} (Academic Press, 1995).

\bibitem{yariv2010a}
A.~Yariv, \emph{Optical electronics in modern communications} (Oxford
  University Press, 1997).
  
\bibitem{rokhsari2004b}
H.~Rokhsari and K.~Vahala, \enquote{{Ultralow loss, high Q, four port resonant
  couplers for quantum optics and photonics},} {PHYSICAL REVIEW LETTERS}
  \textbf{{92}} ({2004}).

\bibitem{yoshiki2012a}
W.~Yoshiki and T.~Tanabe, \enquote{Optical bistable behaviour in side-coupled cavity,} (2012) (preparing for submission).

\bibitem{oxborrow2007a}
M.~Oxborrow, \enquote{{Traceable 2-D finite-element simulation of the
  whispering-gallery modes of axisymmetric electromagnetic resonators},} IEEE
  Trans. Microw. Theory Tech. \textbf{{55}}, {1209--1218} ({2007}).

\bibitem{manolatou1999a}
C.~Manolatou, M.~Khan, S.~Fan, P.~Villeneuve, H.~Haus, and J.~Joannopoulos,
  \enquote{{Coupling of modes analysis of resonant channel add-drop filters},}
  IEEE J. Quantum Electron. \textbf{{35}}, {1322--1331} ({1999}).

\bibitem{miya1979a}
T.~Miya, Y.~Terunuma, T.~Hosaka, and T.~Miyashita, \enquote{{Ultimate low-loss
  single-mode fibre at 1.55 {$\mathrm{\mu}$}m},} Electron. Lett. \textbf{{15}},
  {106--108} ({1979}).

\bibitem{toroid2}
T.~Kippenberg, S.~Spillane, and K.~Vahala, \enquote{{Demonstration of
  ultra-high-Q small mode volume toroid microcavities on a chip},} Appl. Phys.
  Lett. \textbf{{85}}, {6113--6115} ({2004}).

\bibitem{maleki2004a}
A.~Savchenkov, V.~Ilchenko, A.~Matsko, and L.~Maleki, \enquote{{Kilohertz
  optical resonances in dielectric crystal cavities},} Phys. Rev. A
  \textbf{{70}}, 051804 ({2004}).

\bibitem{rokhsari2004a}
H.~Rokhsari, S.~Spillane, and K.~Vahala, \enquote{{Loss characterization in
  microcavities using the thermal bistability effect},} Appl. Phys. Lett.
  \textbf{{85}}, {3029--3031} ({2004}).

\bibitem{spillane2003a}
S.~Spillane, T.~Kippenberg, O.~Painter, and K.~Vahala, \enquote{{Ideality in a
  fiber-taper-coupled microresonator system for application to cavity quantum
  electrodynamics},} Phys. Rev. Lett. \textbf{{91}} ({2003}).

\bibitem{cai2000a}
M.~Cai, O.~Painter, and K.~Vahala, \enquote{{Observation of critical coupling
  in a fiber taper to a silica-microsphere whispering-gallery mode system},}
  Phys. Rev. Lett. \textbf{{85}}, {74--77} ({2000}).

\bibitem{yanik2003a}
M.~Yanik, S.~Fan, and M.~Soljacic, \enquote{{High-contrast all-optical bistable
  switching in photonic crystal microcavities},} Appl. Phys. Lett.
  \textbf{{83}}, {2739--2741} ({2003}).

\bibitem{barclay2005a}
P.~Barclay, K.~Srinivasan, and O.~Painter, \enquote{Nonlinear response of
  silicon photonic crystal microresonators excited via an integrated waveguide
  and fiber taper,} Opt. Express \textbf{{13}}, {801--820} (2005).

\bibitem{bauters2011a}
J.~Bauters, M.~Heck, D.~John, J.~Barton, C.~Bruinink, A.~Leinse, R.~Heideman,
  D.~Blumenthal, and J.~Bowers, \enquote{{Planar waveguides with less than
  0.1~dB/m propagation loss fabricated with wafer bonding},} Opt. Express
  \textbf{{19}}, {24090--24101} (2011).

\bibitem{lee2012a}
H.~Lee, T.~Chen, J.~Li, O.~Painter, and K.~Vahala, \enquote{Ultra-low-loss
  optical delay line on a silicon chip,} Nature Comm. \textbf{{3}}, {867}
  (2012).
\end{thebibliography}

\end{document}